# The relation between black hole spin and molecular gas in massive galaxies

Yongyun Chen (陈永云),[1]★ Qiusheng Gu(顾秋生),[2]★ Junhui Fan(樊军辉),[3] Xiaoling Yu(俞效龄),[1] Nan Ding(丁楠),[4] Xiaotong Guo(郭晓通),[5] and Dingrong Xiong(熊定荣)[6]

[1]*College of Physics and Electronic Engineering, Qujing Normal University, Qujing 655011, P. R. China*
[2]*School of Astronomy and Space Science, Nanjing University, Nanjing 210093, P. R. China*
[3]*Center for Astrophysics, Guang zhou University, Guang zhou 510006, P. R. China*
[4]*School of Physical Science and Technology, Kunming University, Kunming 650214, P. R. China*
[5]*School of mathematics and physics, Anqing Normal University, Anqing 246011, P. R. China*
[6]*Yunnan Observatories, Chinese Academy of Sciences, Kunming 650011, P. R. China*

**ABSTRACT**
Molecular gas is the key probe for the complex interaction between the accretion of black holes and star formation of the host galaxy of active galactic nuclei (AGN). The molecular gas discovered around the AGN indicates that this gas is providing fuel for the AGN. According to the theoretical model of the relativistic jet, the spin of a black hole enhances the relativistic jet of AGN. The spin of the black hole is used as an indicator of AGN activity. Therefore, we study the relationship between the activity of AGN and molecular gas. We find a significant strong correlation between molecular gas fraction and CO luminosity and black hole spin for the early-type galaxies. However, there is no correlation between molecular gas fraction and CO luminosity and black hole spin for the late-type galaxies. These results indicate that the spin of black holes mainly regulates the accretion of molecular gas in massive early-type galaxies. The activity of AGN depends on the amount of gas.

**Key words:** galaxies: active – galaxies: formation – galaxies: general – galaxies: jets.

## 1 INTRODUCTION

The evolution of galaxies is closely related to the growth of their supermassive black holes (SMBHs), which is proved by the correlation between SMBHs mass and the bulge properties of host galaxies (Magorrian et al. 1998; Ferrarese & Merritt 2000; Gebhardt et al. 2000). It is generally believed that the coevolution of SMBHs and host galaxies is regulated by the feedback of active galactic nuclei (AGN; e.g. Kormendy & Ho 2013). In the process of mass accretion on the SMBHs, a large amount of energy is released. A small part of this energy is coupled with the environment around the SMBHs, which can heat and/or remove gas from the host galaxy, thus stopping ongoing star formation activity (e.g. Somerville et al. 2008; Schaye et al. 2015; Sijacki et al. 2015; Dubois et al. 2016; Nelson et al. 2018). AGN feedback can also affect galaxy halo, prevent the condensation of cold gas and warm gas, and further hinder the star formation over long time-scales (Bower et al. 2006; Croton et al. 2006; Fabian 2012; Gaspari, Tombesi & Cappi 2020).

Over the past decade, many efforts have focused on studying whether feedback from AGN can effectively exclude cold gas from host galaxies and prevent star formation. Compared with the control samples and/or the main-sequence of star formation, some studies have found that there is no difference in star formation between inactive galaxies and those with host AGN (e.g. Rosario et al. 2018; Kirkpatrick et al. 2019; Schulze et al. 2019; Koss et al. 2021; Smirnova-Pinchukova et al. 2022), which may imply that the feedback from AGN may not affect star formation or delayed feedback. According to studying the relationship between the luminosity of AGN and molecular gas, star formation rate and star formation efficiency, many authors have found that the luminous AGN with low redshift is mainly hosted in typical star-forming galaxies (e.g. Harrison et al. 2012; Rosario et al. 2012, 2013; Husemann et al. 2017; Bernhard et al. 2019; Grimmett et al. 2020; Yesuf & Ho 2020; Vietri et al. 2022). However, the more luminous active systems often appear in starbursts (e.g. Bernhard et al. 2016; Pitchford et al. 2016; Kirkpatrick et al. 2020). However, it is unclear how the AGN can effectively remove heat gas from the host galaxy and affect ongoing star formation (Harrison 2017).

The relationship between SMBHs accretion and star formation in host galaxies has always been a hot research topic. There has also been no consensus reached in this regard. Some people believe that the star formation rate is closely related to the luminosity of AGN in a wide range of redshifts, which means that SMBHs and its host grow simultaneously (e.g. Bonfield et al. 2011; Rosario et al. 2012; Lanzuisi et al. 2017; Stemo et al. 2020), while others found only moderate or no correlation (Rosario et al. 2013; Azadi et al. 2015; Stanley et al. 2015, 2017). This inconsistency is caused by the different time-scales of the measured SFR and AGN variability (Hickox et al. 2014). Many studies have not found a significant correlation between SFR and AGN luminosity for high-redshift AGN (e.g. Shao et al. 2010; Schulze et al. 2019; Stemo et al. 2020). For low redshift, the correlation is moderate in weak AGN (e.g. Shimizu et al. 2017), but stronger in more powerful systems (e.g. Netzer 2009; Imanishi et al. 2011; Lani, Netzer & Lutz 2017; Zhuang & Ho 2020).

★ E-mails: ynkmcyy@yeah.net (YC); qsgu@nju.edu.cn (QG)

The connection between SMBHs accretion and host galaxy growth occurs naturally because both depend on the same fuel reservoir, which is mainly supplemented by a similar process of driving gas inward (Kormendy & Kennicutt 2004). A large amount of molecular gas ($\geq 10^8$ M$_\odot$) has been found in elliptical galaxies at the centre of nearby clusters, indicating that the cold gas is feeding the AGN (e.g. Edge 2001; Salomé & Combes 2003; David et al. 2014; Vantyghem et al. 2014; Russell et al. 2017, 2019; Olivares et al. 2019; Rose et al. 2020; North et al. 2021). Russell et al. (2019) studied the relation between X-ray cavity power and molecular gas mass and found that there is a correlation between X-ray cavity power and molecular gas mass for 12 central cluster galaxies (their fig. 7). Shangguan et al. (2020) found that there is a correlation between the AGN luminosity and CO(2–1) luminosity and black hole mass for 40 $z < 0.3$ Palomar–Green quasars, which implies that AGN activity is related to the cold gas reservoir of the host. Fujita et al. (2023) found that there is a correlation between AGN jet power and molecular gas mass for massive elliptical galaxies. However, some studies have found that the luminosity of AGN is not related to molecular gas. Molina et al. (2023) studied the relationship between AGN properties and molecular gas of the host galaxy by using 138 local type 1 AGNs with low redshifts ($z \leq 0.5$) and found that there is no correlation between the CO(2–1) luminosity and AGN luminosity. Theoretical studies (Shlosman, Frank & Begelman 1989) and numerical simulations (Pizzolato & Soker 2005, 2010; Wagner, Bicknell & Umemura 2012; McNamara et al. 2016) have shown that cold gases can play a fundamental role in fueling nearby AGN. Although there are some studies on the relationship between the activity of AGN and molecular gas, it has been uncertain whether the activity of AGN depends on the accretion of molecular gas.

According to the theory of jet formation, the jet power of AGN mainly depends on the black hole spin (Blandford & Znajek 1977). Some observations also show that the black hole activity depends on the spin of the black hole (e.g. Narayan & McClintock 2012; Steiner, McClintock & Narayan 2013; Ünal & Loeb 2020; Chen et al. 2021). Therefore, the black hole spin can be used as an indicator of AGN activity. In this article, we use black hole spin to investigate whether the activity of AGN affects the accretion of molecular gas. We therefore mainly study the relationship between black hole spin and molecular gas for AGN. The second part describes the sample; the third part presents the results and discussion; the fourth part is the conclusion.

## 2 THE SAMPLE

### 2.1 The sample of massive galaxies

We select an AGN sample with reliable molecular gas mass, black hole mass, bolometric luminosity, and redshift. First, we consider a large sample of AGN with reliable molecular gas mass, star formation rate, and stellar mass. We use the catalogue of Koss et al. (2021). Koss et al. (2021) presented the host-galaxy molecular gas properties of a sample of 213 nearby AGN galaxies drawn from the 70-month catalogue of *Swift*'s Burst Alert Telescope (BAT). They used the CO(2–1) line to get the molecular gas mass and CO luminosity. Secondly, we consider that AGN has reliable black hole mass and bolometric luminosity. We use the catalogue of Koss et al. (2022). Koss et al. (2022) presented the AGN catalogue and optical spectroscopy for the second data release of the *Swift*–BAT AGN Spectroscopic Survey (BASS DR2). Koss et al. (2022) estimated the bolometric luminosity using the intrinsic luminosity in the 14–150 keV range. They used a 14–150 keV bolometric correction of 8 based on the factor of 20 for the 2–10 keV range (Vasudevan & Fabian 2009). Finally, in order to calculate the spin of the black hole, we consider the AGN with 1.4 GHz radio flux. The 1.4 GHz radio flux comes from the NED. In total, we get 68 AGNs. We use the morphological parameter $T$ to separate AGN into early-type ($T < 0$) and late-type galaxies ($T > 0$). The morphology indicator $T$-type is obtained from the HyperLEDA data base (Makarov et al. 2014). We divide these AGNs into FR I and FR II radio sources based on radio luminosity of 1.4 GHz (e.g. Fanaroff & Riley 1974; Urry & Padovani 1995; Ledlow & Owen 1996; Mingo et al. 2019). The FR I radio source has a low radio luminosity of 1.4 GHz ($L_{1.4\text{GHz}} < 10^{24.5}$ WHz$^{-1}$), while the FR II radio source has a relatively high radio luminosity of 1.4 GHz ($L_{1.4\text{GHz}} > 10^{24.5}$ WHz$^{-1}$).

### 2.2 The spin of black hole

Daly (2019) got that the spin function $f(j)$ of a black hole can be normalized by its maximum value $f_{\max}$,

$$\frac{f(j)}{f_{\max}} = \left(\frac{L_j}{g_j L_{\text{Edd}}}\right) \left(\frac{L_{\text{bol}}}{g_{\text{bol}} L_{\text{Edd}}}\right)^{-A}, \quad (1)$$

where $L_j$ is the beam power, $g_j = 0.1$, and $g_{\text{bol}} = 1$ are adopted (Daly 2019). Daly, Stout & Mysliwiec (2018) defined the $A = a/(a + b)$, $a$ and $b$ are the coefficients of the Fundamental Plane of black hole activity, where $A = 0.43$ (Merloni, Heinz & di Matteo 2003). The black hole spin can be calculated by using the spin function, $\sqrt{f(j)/f_{\max}} = j(1 + \sqrt{1 - j^2})^{-1}$. For $f(j)/f_{\max} \leq 1$, this implies (Daly 2019)

$$j = \frac{2\sqrt{f(j)/f_{\max}}}{f(j)/f_{\max} + 1}. \quad (2)$$

The black spin is calculated using equation (2) and values of $f(j)/f_{\max} > 1$ are set equal to 1 (Daly 2019). The values of black hole spin are shown in Table 1. The black hole spin obtained by this method can be compared with the black hole spin obtained by other methods, such as the X-ray reflection method (Reynolds 2014). The spin of GX 339–4 is $0.94 \pm 0.02$ using the X-ray reflection method (Miller et al. 2009), and the spin is $0.92 \pm 0.06$ using the method of Daly (2019). The spin obtained by the method of Daly (2019) is consistent with that obtained by the X-ray reflection method. Due to the limitations of observation, it is very difficult to directly measure the black hole spin of large samples based on the X-ray reflection method. Therefore, we use the method of Daly (2019) to measure the spin of our sample.

### 2.3 The beam power

The beam power can be estimated by the following formula (Cavagnolo et al. 2010),

$$L_j \approx 5.8 \times 10^{43} \left(\frac{L_{\text{radio}}}{10^{40} \text{erg s}^{-1}}\right)^{0.70} \text{ erg s}^{-1}, \quad (3)$$

where $L_{\text{radio}}$ is the radio luminosity at 1.4 GHz in units of erg s$^{-1}$. The 1.4 GHz radio luminosity is estimated by using the formula $L_\nu = 4\pi d_L^2 S_\nu$, $d_L(z) = \frac{c}{H_0}(1+z)\int_0^z[\Omega_\Lambda + \Omega_m(1+z')^3]^{-1/2}dz'$, where $d_L$ is the luminosity distance (Venters, Pavlidou & Reyes 2009). We make a K-correction for the 1.4 GHz radio flux using $S_\nu = S_\nu^{\text{obs}}(1+z)^{\alpha-1}$ and $\alpha = 0$ (e.g. Abdo et al. 2010; Komossa, Xu & Wagner 2018). Equation (3) can be used to estimate the beam power of AGN, including both FR I and FR II sources (e.g. Meyer et al. 2011; Heckman & Best 2014; Nisbet & Best 2016; Sabater et al. 2019).

Table 1. The sample of massive galaxies.

| Name (1) | RA (2) | DEC (3) | Redshift (4) | T-type (5) | log $M_*$ (6) | SFR (7) | log $L_{CO}$ (8) | log $M_{H_2}$ (9) | log $M_{BH}$ (10) | log $L_{bol}$ (11) | $S_\nu$ (12) | log $L_{radio}$ (13) | log $L_j$ (14) | $f_j/f_{max}$ (15) | $j$ (16) | Class (17) |
|---|---|---|---|---|---|---|---|---|---|---|---|---|---|---|---|---|
| NGC235A | 10.71986 | −23.54019 | 0.022 | −2.2 | 10.9 | 6.5 | 9.06 | 9.7 | 8.49 | 44.61 | 0.0422 | 38.81 | 42.93 | 0.015 | 0.243 | FRI |
| ESO195-IG021NED03 | 15.14593 | −47.86886 | 0.048 | −1.86 | 10.95 | 6.36 | 9.36 | 10.03 | 8.06 | 44.8 | 0.0357 | 39.42 | 43.36 | 0.060 | 0.461 | FRI |
| NGC424 | 17.86474 | −38.08319 | 0.011 | 0.1 | 10.49 | 0.6 | 7.78 | 8.5 | 7.49 | 44.22 | 0.023 | 37.93 | 42.31 | 0.020 | 0.278 | FRI |
| IC1657 | 18.52913 | −32.65182 | 0.012 | 3.8 | 10.62 | 3.18 | 8.67 | 9.39 | 7.68 | 43.51 | 0.0344 | 38.16 | 42.48 | 0.046 | 0.411 | FRI |
| NGC513 | 21.11072 | 33.79914 | 0.019 | 5.5 | 10.86 | 7.17 | 9.66 | 9.61 | 7.63 | 44.05 | 0.0529 | 38.78 | 42.91 | 0.078 | 0.519 | FRI |
| Z459-58 | 22.10173 | 16.45963 | 0.038 | 0 | 10.83 | 9.87 | 9.17 | 9.77 | 8.1 | 44.5 | 0.0136 | 38.8 | 42.92 | 0.028 | 0.324 | FRI |
| ESO353-9 | 22.96012 | −33.11902 | 0.016 | 3.7 | 10.68 | 5.22 | 9.05 | 9.64 | 7.96 | 43.82 | 0.0274 | 38.36 | 42.61 | 0.032 | 0.349 | FRI |
| UGC1479 | 30.07942 | 24.47393 | 0.017 | 3.5 | 10.6 | 2.9 | 8.72 | 9.32 | 7.51 | 43.81 | 0.012 | 38.03 | 42.38 | 0.035 | 0.359 | FRI |
| Mrk590 | 33.63951 | −0.76673 | 0.026 | 1 | 11.2 | 3.65 | 9 | 9.56 | 7.57 | 44.24 | 0.0162 | 38.55 | 42.75 | 0.048 | 0.420 | FRI |
| IC1816 | 37.9629 | −36.67242 | 0.017 | 2.1 | 10.74 | 3.38 | 8.89 | 9.51 | 7.37 | 44.01 | 0.0336 | 38.46 | 42.69 | 0.069 | 0.490 | FRI |
| NGC1125 | 42.91859 | −16.6508 | 0.011 | 0 | 10.43 | 1.7 | 9.26 | 8.82 | 7.39 | 43.81 | 0.058 | 38.35 | 42.61 | 0.068 | 0.487 | FRI |
| MCG-2-8-38 | 45.01781 | −10.82518 | 0.033 | 2.5 | 10.83 | 1.03 | 8.58 | 9.25 | 7.97 | 44.49 | 0.0175 | 38.78 | 42.91 | 0.032 | 0.348 | FRI |
| NGC1229 | 47.04468 | −22.95974 | 0.036 | 3.5 | 10.75 | 17.22 | 9.12 | 9.77 | 8.32 | 44.99 | 0.0194 | 38.89 | 42.98 | 0.015 | 0.240 | FRI |
| UGC2638 | 49.25902 | 1.25553 | 0.024 | 2 | 10.56 | 12.48 | 9.03 | 9.76 | 7.51 | 44.07 | 0.027 | 38.68 | 42.84 | 0.076 | 0.513 | FRI |
| ESO549-49 | 60.60697 | −18.04731 | 0.026 | 3.8 | 10.92 | 14.92 | 9.45 | 10.12 | 8.16 | 44.42 | 0.0248 | 38.73 | 42.87 | 0.025 | 0.308 | FRI |
| 3C120 | 68.29581 | 5.35439 | 0.033 | −1.7 | 11 | 1.69 | 8.61 | 9.56 | 7.74 | 45.22 | 2.95 | 41.01 | 44.47 | 0.257 | 0.807 | FRII |
| Mrk618 | 69.0917 | −10.37589 | 0.035 | 3 | 10.92 | 20.86 | 9.89 | 10.1 | 7.72 | 44.51 | 0.017 | 38.81 | 42.93 | 0.046 | 0.410 | FRI |
| LEDA86269 | 71.03769 | 28.21668 | 0.011 | 3.2 | 10.63 | 1.16 | 9.6 | 9.38 | 7.98 | 43.92 | 0.0286 | 37.99 | 42.36 | 0.016 | 0.248 | FRI |
| LEDA75258 | 75.53757 | 3.53047 | 0.016 | 0 | 10.07 | 0.1 | 7.7 | 8.35 | 7.02 | 43.74 | 0.0187 | 38.17 | 42.48 | 0.088 | 0.546 | FRI |
| MCG-2-15-4 | 83.45886 | −13.35483 | 0.029 | −1.31 | 10.68 | 7.43 | 8.46 | 9.1 | 8.1 | 44.47 | 0.0508 | 39.12 | 43.15 | 0.048 | 0.419 | FRI |
| LEDA17883 | 87.1184 | −47.75901 | 0.05 | 1 | 11.03 | 62.7 | 8.66 | 9.3 | 8.19 | 44.55 | 0.0735 | 39.77 | 43.60 | 0.113 | 0.604 | FRI |
| UGC3374 | 88.72394 | 46.43932 | 0.02 | 3.6 | 11.03 | 3.93 | 9.85 | 9.58 | 6.61 | 44.93 | 0.246 | 39.49 | 43.41 | 0.394 | 0.901 | FRI |
| H0557-385 | 89.5082 | −38.3347 | 0.033 | 0 | 10.98 | 1.16 | 8.61 | 9.25 | 7.4 | 44.7 | 0.0346 | 39.07 | 43.12 | 0.090 | 0.549 | FRI |
| VIIZw73 | 97.60729 | 63.67782 | 0.04 | 0 | 10.76 | 18.26 | 9.71 | 9.72 | 7.18 | 44.65 | 0.0119 | 38.79 | 42.91 | 0.079 | 0.521 | FRI |
| ESO490-26 | 100.04874 | −25.89584 | 0.025 | 1.8 | 10.94 | 3.15 | 9.2 | 9.84 | 7.15 | 44.57 | 0.0385 | 38.88 | 42.98 | 0.104 | 0.584 | FRI |
| LEDA96373 | 111.6096 | −35.90678 | 0.03 | −1 | 11.25 | 1.6 | 8.18 | 9.17 | 9.62 | 44.5 | 0.172 | 39.68 | 43.54 | 0.016 | 0.247 | FRI |
| Z118-36 | 119.9729 | 23.39024 | 0.029 | 2.5 | 11.03 | 15.85 | 9.85 | 10.04 | 8.43 | 44.61 | 0.018 | 38.67 | 42.84 | 0.013 | 0.227 | FRI |
| ESO209-12 | 120.49101 | −49.77796 | 0.04 | 1 | 11.31 | 10.68 | 9.3 | 9.96 | 7.82 | 44.74 | 0.0578 | 39.47 | 43.39 | 0.093 | 0.558 | FRI |
| Mrk1210 | 121.02439 | 5.11381 | 0.014 | 1.03 | 10.24 | 0.9 | 8.15 | 8.81 | 6.86 | 44.25 | 0.114 | 38.81 | 42.93 | 0.186 | 0.727 | FRI |
| Fairall272 | 125.75451 | −4.93522 | 0.022 | 1.1 | 10.4 | 2.52 | 8.52 | 9.14 | 8.11 | 44.63 | 0.0372 | 38.76 | 42.89 | 0.023 | 0.295 | FRI |
| Mrk18 | 135.4931 | 60.15201 | 0.011 | 1.5 | 10.05 | 1.4 | 9.02 | 8.57 | 7.72 | 43.36 | 0.0263 | 38 | 42.36 | 0.039 | 0.381 | FRI |
| MCG-1-24-12 | 140.1926 | −8.05592 | 0.02 | 5 | 10.21 | 1.14 | 8.62 | 9.26 | 7.66 | 44.4 | 0.0262 | 38.5 | 42.71 | 0.034 | 0.355 | FRI |
| ESO565-19 | 143.68213 | −21.92761 | 0.016 | −3.6 | 10.89 | 4.28 | 8.56 | 9.19 | 8.45 | 44.56 | 0.0437 | 38.54 | 42.74 | 0.011 | 0.207 | FRI |
| UGC5101 | 143.96469 | 61.35337 | 0.039 | −0.5 | 10.85 | 128.89 | 9.79 | 10.42 | 8.19 | 44.96 | 0.147 | 39.85 | 43.66 | 0.086 | 0.541 | FRI |
| ESO263-13 | 152.44971 | −42.8111 | 0.034 | 2.9 | 10.88 | 0.48 | 8.48 | 9.05 | 8.06 | 44.86 | 0.0154 | 38.74 | 42.88 | 0.019 | 0.268 | FRI |
| Fairall1149 | 153.33189 | −35.98268 | 0.028 | 1.2 | 11.31 | 1.2 | 8.08 | 8.71 | 8.16 | 44.45 | 0.0785 | 39.29 | 43.27 | 0.060 | 0.462 | FRI |
| ESO317-41 | 157.8456 | −42.06023 | 0.019 | 4.4 | 10.75 | 11.84 | 9.07 | 9.7 | 7.77 | 44.15 | 0.0636 | 38.87 | 42.97 | 0.068 | 0.489 | FRI |
| LEDA93974 | 160.09206 | −46.42426 | 0.024 | −2 | 10.7 | 3.88 | 8.58 | 9.24 | 7.9 | 44.31 | 0.0393 | 38.84 | 42.95 | 0.047 | 0.414 | FRI |
| ESO438-9 | 167.70047 | −28.5008 | 0.024 | 2.1 | 10.6 | 12.19 | 9.04 | 9.67 | 6.87 | 43.77 | 0.0152 | 38.43 | 42.66 | 0.159 | 0.689 | FRI |
| Mrk732 | 168.45702 | 9.58614 | 0.029 | −3.2 | 10.91 | 15.04 | 9.89 | 10.16 | 6.59 | 44.17 | 0.0376 | 39 | 43.07 | 0.390 | 0.899 | FRI |
| NGC3822 | 175.54633 | 10.27762 | 0.019 | −1.4 | 11 | 14.05 | 9.27 | 9.9 | 7.43 | 43.91 | 0.0268 | 38.5 | 42.71 | 0.075 | 0.508 | FRI |
| IC751 | 179.71877 | 42.5702 | 0.031 | 2.9 | 10.96 | 10.21 | 9.71 | 9.59 | 8.57 | 44.38 | 0.043 | 39.12 | 43.15 | 0.028 | 0.327 | FRI |
| UGC7064 | 181.18063 | 31.17703 | 0.025 | 3.1 | 10.91 | 7.73 | 9.8 | 9.93 | 7.57 | 44 | 0.0167 | 38.52 | 42.72 | 0.058 | 0.456 | FRI |
| NGC4253 | 184.61073 | 29.81284 | 0.013 | 1 | 10.42 | 3.02 | 9.32 | 8.99 | 6.82 | 43.75 | 0.0403 | 38.32 | 42.59 | 0.145 | 0.666 | FRI |
| NGC4507 | 188.90138 | −39.90903 | 0.012 | 1.8 | 10.81 | 1.97 | 8.59 | 9.21 | 7.81 | 44.8 | 0.0661 | 38.45 | 42.68 | 0.017 | 0.258 | FRI |
| ESO506-27 | 189.72795 | −27.30739 | 0.024 | −0.8 | 11 | 2.45 | 8.62 | 9.32 | 8.84 | 45.07 | 0.0737 | 39.13 | 43.15 | 0.010 | 0.200 | FRI |

**Table 1** – *continued*

| Name (1) | RA (2) | DEC (3) | Redshift (4) | T-type (5) | log $M_*$ (6) | SFR (7) | log $L_{CO}$ (8) | log $M_{H_2}$ (9) | log $M_{BH}$ (10) | log $L_{bol}$ (11) | $S_\nu$ (12) | log $L_{radio}$ (13) | log $L_j$ (14) | $f_j/f_{max}$ (15) | $j$ (16) | Class (17) |
|---|---|---|---|---|---|---|---|---|---|---|---|---|---|---|---|---|
| ESO323-77 | 196.6087 | −40.41484 | 0.016 | −2 | 10.93 | 8.66 | 8.98 | 9.63 | 6.52 | 44.12 | 0.032 | 38.39 | 42.63 | 0.166 | 0.699 | FRI |
| MCG-3-34-64 | 200.60159 | −16.72865 | 0.017 | −2 | 10.78 | 7.1 | 9.47 | 9.19 | 8.37 | 44.27 | 0.275 | 39.38 | 43.33 | 0.063 | 0.471 | FRI |
| NGC5252 | 204.56648 | 4.54302 | 0.023 | −2 | 11.08 | 1.14 | 8.15 | 8.8 | 9 | 44.95 | 0.012 | 38.3 | 42.57 | 0.002 | 0.098 | FRI |
| IC4329A | 207.3304 | −30.30936 | 0.016 | −1 | 10.9 | 1.16 | 7.85 | 8.53 | 7.65 | 45.06 | 0.0664 | 38.72 | 42.87 | 0.026 | 0.312 | FRI |
| NGC5548 | 214.49802 | 25.13813 | 0.017 | 0.4 | 10.84 | 1.73 | 8.68 | 9.35 | 7.72 | 44.52 | 0.0282 | 38.39 | 42.63 | 0.023 | 0.298 | FRI |
| IRAS15091-2107 | 227.99947 | −21.31739 | 0.044 | 1.5 | 11.1 | 15.49 | 9.46 | 10.11 | 6.94 | 44.98 | 0.0469 | 39.47 | 43.39 | 0.233 | 0.783 | FRI |
| MCG-1-40-1 | 233.33632 | −8.70076 | 0.023 | 3.2 | 10.88 | 3.56 | 9.79 | 9.7 | 7.63 | 44.44 | 0.213 | 39.55 | 43.45 | 0.183 | 0.723 | FRI |
| NGC5995 | 237.10361 | −13.75793 | 0.025 | 0.9 | 11.17 | 14.99 | 9.9 | 10.09 | 7.77 | 44.62 | 0.03 | 38.76 | 42.90 | 0.036 | 0.366 | FRI |
| NGC6240 | 253.24486 | 2.40087 | 0.025 | 0 | 11.32 | 75.65 | 10.11 | 10.65 | 9.2 | 45.04 | 0.426 | 39.91 | 43.70 | 0.023 | 0.297 | FRI |
| LEDA90334 | 294.38757 | −6.2179 | 0.01 | −5 | 10.27 | 2.37 | 9.46 | 9.15 | 6.61 | 43.61 | 0.0422 | 38.15 | 42.47 | 0.167 | 0.700 | FRI |
| Z373-11 | 306.73297 | −2.27736 | 0.029 | −2.8 | 10.12 | 4.85 | 8.97 | 9.49 | 7.42 | 44.48 | 0.0297 | 38.9 | 42.99 | 0.082 | 0.528 | FRI |
| NGC6921 | 307.11975 | 25.72294 | 0.014 | 7.9 | 10.48 | 11.06 | 9.67 | 9.62 | 8.72 | 43.94 | 0.0848 | 38.72 | 42.86 | 0.019 | 0.270 | FRI |
| Mrk509 | 311.04108 | −10.7237 | 0.035 | 1.23 | 11.11 | 9.14 | 9.02 | 9.71 | 8.05 | 45.27 | 0.0186 | 38.85 | 42.96 | 0.015 | 0.241 | FRI |
| H2106-099 | 317.29083 | −9.67102 | 0.027 | −2 | 10.74 | 0.59 | 8.15 | 8.79 | 7.51 | 44.35 | 0.0053 | 38.07 | 42.42 | 0.022 | 0.289 | FRI |
| NGC7130 | 327.08142 | −34.95125 | 0.016 | 1.2 | 10.9 | 33.32 | 9.44 | 10.07 | 7.04 | 43.81 | 0.19 | 39.17 | 43.18 | 0.404 | 0.905 | FRI |
| UGC11910 | 331.75867 | 10.23348 | 0.027 | 3.4 | 11.01 | 19.58 | 9.21 | 9.79 | 8.15 | 44.84 | 0.103 | 39.36 | 43.31 | 0.046 | 0.409 | FRI |
| Mrk915 | 339.1947 | −12.54436 | 0.024 | 7 | 10.82 | 2.98 | 8.28 | 9.07 | 7.93 | 44.45 | 0.07 | 39.1 | 43.13 | 0.059 | 0.460 | FRI |
| UGC12138 | 340.07114 | 8.0536 | 0.025 | 1 | 10.67 | 2.39 | 9.03 | 9.7 | 7.08 | 44.21 | 0.00827 | 38.21 | 42.51 | 0.055 | 0.443 | FRI |
| NGC7469 | 345.81503 | 8.87378 | 0.016 | 1.1 | 11.04 | 35.56 | 9.58 | 10.21 | 6.96 | 44.41 | 0.146 | 39.06 | 43.11 | 0.209 | 0.756 | FRI |
| Mrk926 | 346.18155 | −8.68567 | 0.048 | 3.5 | 11.23 | 7.11 | 8.84 | 9.7 | 7.98 | 45.62 | 0.0225 | 39.21 | 43.21 | 0.021 | 0.283 | FRI |
| NGC7603 | 349.73615 | 0.24422 | 0.029 | 3 | 11.37 | 9.03 | 9.84 | 9.9 | 8.59 | 44.82 | 0.0211 | 38.74 | 42.88 | 0.010 | 0.195 | FRI |
| NGC7682 | 352.26663 | 3.53329 | 0.017 | 1.6 | 10.64 | 14.89 | 8.3 | 8.99 | 7.84 | 44.41 | 0.054 | 38.69 | 42.85 | 0.036 | 0.366 | FRI |

*Note.* Column (1) is the name of sources; column (2) is the right ascension in decimal degrees; column (3) is (delineation) in decimal degrees; column (4) is the redshift; column (5) is the morphology parameter; column (6) is the stellar mass; column (7) is the star formation rate; column (8) is the CO(2−1) luminosity, in units K km s$^{-1}$ pc$^2$; column (9) is the molecular gas mass; column (10) is the black hole mass; column (11) is the bolometric luminosity, in units erg s$^{-1}$; column (12) is the 1.4 GHz radio flux, in units Jy; column (13) is the 1.4 GHz radio luminosity, in units erg s$^{-1}$; column (14) is the beam power, in units erg s$^{-1}$; column (15) is the spin function; column (16) is the back hole spin; column (17) is the class of sources (FRI is FR I radio galaxies, FRII is FR II radio galaxies).

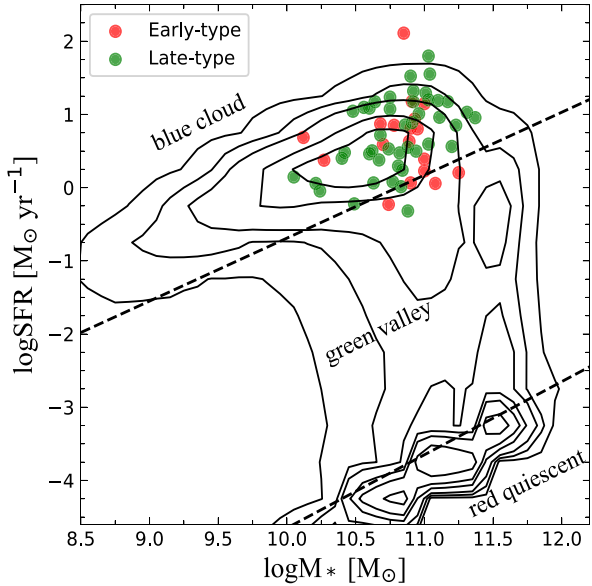
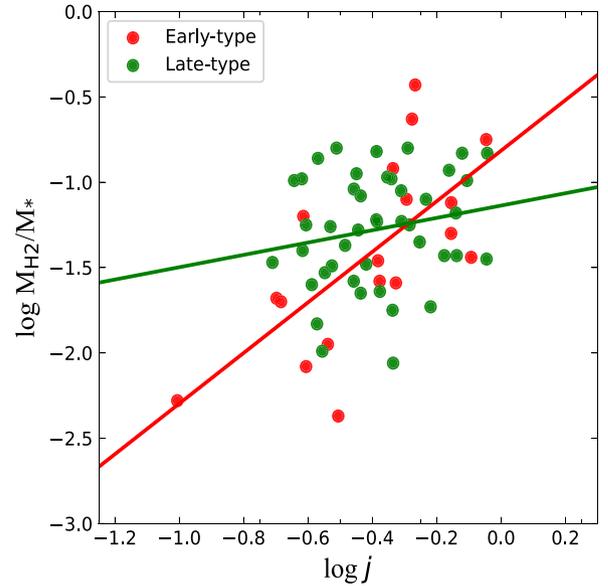

**Figure 1.** Relation between star formation rate and stellar mass for AGN. The red dot is early-type galaxies, and the green dot is late-type galaxies. The contour is the sample of Chang et al. (2015). The two dashed lines divide galaxies into blue galaxies (star-forming) and red galaxies with the green valley galaxies located between them.

**Figure 2.** Relation between spin of black hole mass and molecular gas fraction for AGN. The red dot is early-type galaxies and the green dot is late-type galaxies. The red line is a linear fit for early-type galaxies. The green line is a linear fit for late-type galaxies.

A $\Lambda$CDM cosmology with $H_0 = 70$ km s$^{-1}$Mpc$^{-1}$, $\Omega_\Lambda = 0.73$, and $\Omega_m = 0.27$ is adopted.

## 3 RESULT AND DISCUSSION

### 3.1 Relation between star formation rate and stellar mass

We plot these AGNs on the relation panel between star formation rate and stellar mass. Fig. 1 shows the relation between star formation rate and stellar mass for AGN. Two dashed lines define blue cloud/star-forming galaxies and red galaxies. The star-forming galaxies are defined as $\log SFR > 0.86 \times \log M_* - 9.29$, the red galaxies have $\log SFR < \log M_* - 14.65$, and the green valley galaxies located between them. The red dot is early-type galaxies and the green dot is late-type galaxies. We find that almost all of these AGN with molecular gas are star-forming galaxies. These AGNs have high stellar mass ($9.5 < \log M_* < 11.5$).

### 3.2 Relation between molecular gas fraction and spin of black hole

The properties of molecular gas in the host galaxy of AGN provide important clues to the role of black hole accretion in the evolution of galaxies. Fig. 2 shows the relation between molecular gas fraction ($M_{H_2}/M_*$) and the spin of a black hole for AGN. We study the relationship between molecular gas fraction and the spin of a black hole through the linear fitting method. In the entire paper, if the $p$-value (significant level) is $p \leq 0.01$, we consider the correlation to be significant. If $p \approx 0.01$–0.05, we consider the correlation to be moderately significant. If $p \approx 0.06$–0.10, we consider the correlation to be weak. If $p > 0.10$, we consider that there is no correlation. There is a significantly strong correlation between molecular gas fraction and spin of a black hole for the whole sample ($r = 0.36$, $p =$ 0.002),

$$\log M_{H_2}/M_* = 0.76(\pm 0.25) \log j - 1.01(\pm 0.11). \quad (4)$$

We also study the relationship between the spin of black holes and molecular gas fraction for early-type and late-type galaxies, respectively. There is also a significant strong correlation between the spin of black hole and molecular gas fraction for early-type galaxies ($r = 0.67$, $p = 0.002$),

$$\log M_{H_2}/M_* = 1.48(\pm 0.41) \log j - 0.81(\pm 0.20). \quad (5)$$

However, there is no correlation between black hole spin and molecular gas fraction for late-type galaxies ($r = 0.19$, $p = 0.22$),

$$\log M_{H_2}/M_* = 0.36(\pm 0.29) \log j - 1.14(\pm 0.12). \quad (6)$$

These results suggest that the black hole spin regulates the molecular gas accretion of massive galaxies. The feedback of AGN may relate to the morphology of galaxies. The AGN feedback has a significant impact on molecular gas for early-type galaxies, while it has little effect on molecular gas for late-type galaxies. Recently, high-resolution numerical simulations of AGN feedback in disc-dominated systems have found that AGN has almost no effect (Gabor & Bournaud 2014). Russell et al. (2019) found a significant correlation between the jet power of AGN and molecular gas mass by using 12 central cluster galaxies. Shangguan et al. (2020) found a weak correlation between molecular gas fraction and AGN continuum luminosity ($\lambda L_\lambda 5100$ Å) by using 40 low redshift ($z < 0.3$) Palomar–Green quasars. Zhuang & Ho (2020) also find a significant correlation between the molecular gas and AGN luminosity. Fujita et al. (2023) found a significant correlation between the jet power of AGN and molecular gas mass for nine massive elliptical galaxies. Babyk et al. (2019) found a weak correlation between the total molecular gas mass and AGN jet power using a sample of nearby early-type galaxies, which provides further evidence that there could be a close connection between the AGN activity and molecular gas. Ward et al.

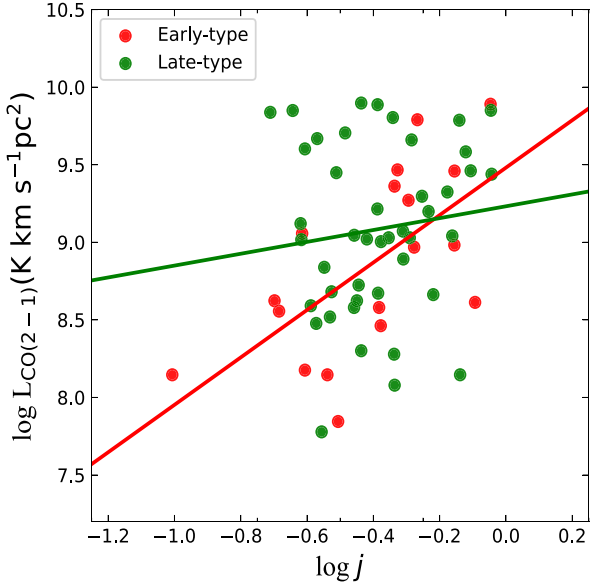

**Figure 3.** Relation between the spin of a black hole and CO(2-1) luminosity for AGN. The red dot is early-type galaxies and the green dot is late-type galaxies. The red line is a linear fit for early-type galaxies. The green line is a linear fit for late-type galaxies.

(2022) found that AGN is preferentially located in galaxies with high-molecular gas fractions.

### 3.3 Relation between CO luminosity and spin of black hole

The relation between CO luminosity and the spin of a black hole for AGN is shown in Fig. 3. There is a moderately significant correlation between black hole spin and CO luminosity for the whole sample ($r = 0.25$, $p = 0.04$),

$$\log L_{CO(2-1)} = 0.77(\pm 0.37) \log j + 9.33(\pm 0.16). \quad (7)$$

There is a significant strong correlation between black hole spin and CO luminosity for early-type galaxies ($r = 0.63$, $p = 0.004$),

$$\log L_{CO(2-1)} = 1.53(\pm 0.47) \log j + 9.48(\pm 0.22). \quad (8)$$

However, there is no correlation between black hole spin and CO luminosity for late-type galaxies ($r = 0.12$, $p = 0.44$),

$$\log L_{CO(2-1)} = 0.38(\pm 0.49) \log j + 9.23(\pm 0.21). \quad (9)$$

These results further indicate that the black hole spin mainly regulates the cold gas accretion for massive galaxies and early-type galaxies. There is a close connection between AGN activity and cold gas supply. Shangguan et al. (2020) found a significant correlation between CO luminosity and $\lambda L_\lambda 5100$ Å by using 40 low redshift ($z < 0.3$) Palomar–Green quasars.

### 3.4 Relation between depletion time and spin of black hole

Fig. 4 shows the relationship between gas depletion time ($t_{dep} \equiv M_{H_2}/SFR$) and spin of a black hole (left panel) and specific star formation rate (right panel) for AGN. There is no correlation between gas depletion time and the spin of the black hole for the whole sample ($r = -0.04$, $p = 0.73$),

$$\log t_{dep} = -0.07(\pm 0.22) \log j - 0.23(\pm 0.09). \quad (10)$$

There is also no correlation between gas depletion time and black hole spin for early-type galaxies ($r = 0.08$, $p = 0.73$),

$$\log t_{dep} = 0.11(\pm 0.30) \log j - 0.20(\pm 0.14). \quad (11)$$

There is no correlation between gas depletion time and black hole spin for late-type galaxies ($r = -0.14$, $p = 0.35$),

$$\log t_{dep} = -0.30(\pm 0.31) \log j - 0.31(\pm 0.13). \quad (12)$$

Molina et al. (2023) studied the relation between gas depletion time and $\lambda L_\lambda 5100$ Å by using 138 local type 1 AGN and found that there is no correlation between gas depletion time and $\lambda L_\lambda 5100$ Å. These results may indicate that the gas depletion time is not related to the activity of AGN. The gas depletion time does not depend on the spin of the black hole but on other physical parameters, such as specific star formation rate (sSFR). Huang & Kauffmann (2014) studied the relationship between the gas depletion time and the physical parameters of the galaxies, such as stellar surface density, concentration index, star formation rate, stellar mass, and sSFR, and found that the primary global parameter correlation is between gas depletion time and specific star formation rate. All other remaining correlations can be proven to be caused by this primary dependency. We also study the relation between gas depletion time and specific star formation rate for our sample. Fig. 4 shows the relation between gas depletion time and specific star formation rate for AGN. There is a significantly strong correlation between gas depletion time and specific star formation rate for the whole sample ($r = -0.61$, $p = 4.15 \times 10^{-8}$). However, there is a moderately strong correlation between gas depletion time and specific star formation rate for early-type galaxies ($r = -0.44$, $p = 0.05$). There is a significantly strong correlation between gas depletion time and specific star formation rate for late-type galaxies ($r = -0.68$, $p = 4.01 \times 10^{-7}$). Our results are consistent with the result of Huang & Kauffmann (2014).

## 4 CONCLUSIONS

We study the relationship between black hole spin and molecular gas using 68 AGN. We estimate the spin parameters of the black hole using radio luminosity, bolometric luminosity, and black hole mass. Our main results are as follows:

(1) There is a significant strong correlation between molecular gas fraction and black hole spin for the whole sample and early-type galaxies. However, there is no correlation between black hole spin and molecular gas fraction and CO luminosity for late-type galaxies. There is a moderately significant correlation between black hole spin and CO luminosity for the whole sample. These results may imply that the black hole spin mainly regulates the molecular gas accretion of massive early-type galaxies.

(2) There is no correlation between gas depletion time and black hole spin for the whole sample, early-type galaxies, and late-type galaxies. These results indicate that the gas depletion time does not depend on the spin of the black hole.

(3) There is a significantly strong correlation between gas depletion time and specific star formation rate for the whole sample and late-type galaxies. However, there is a moderately strong correlation between gas depletion time and specific star formation rate for early-type galaxies. These results show that the gas depletion time depends on the specific star formation rate.

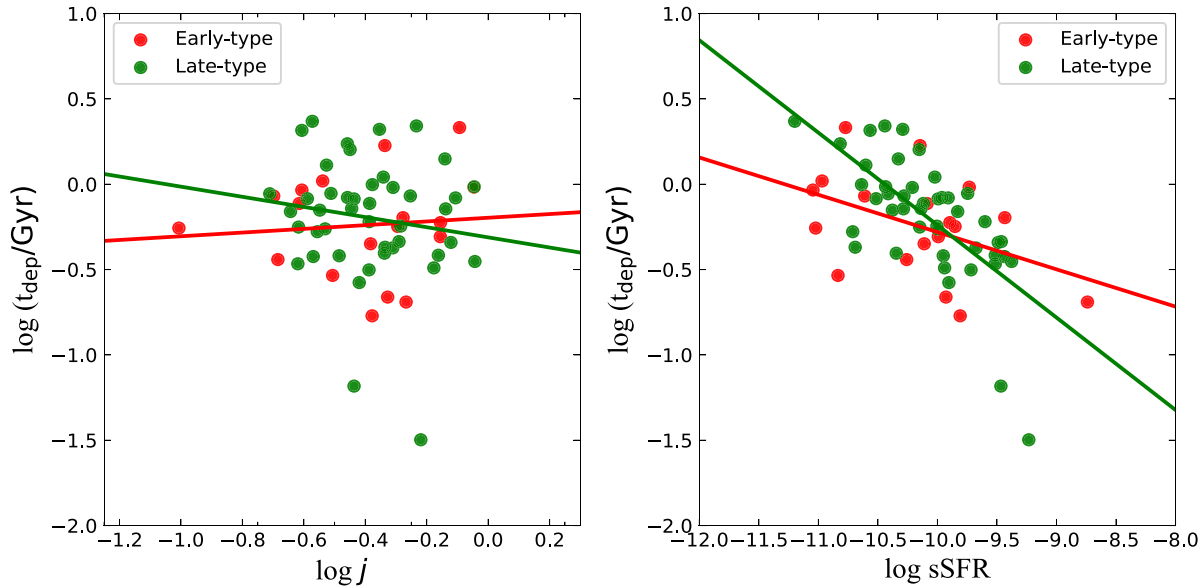

**Figure 4.** Relation between the spin of a black hole (left panel) and specific star formation rate (right panel) and gas depletion time for AGN. The red dot is early-type galaxies and the green dot is late-type galaxies. The red line is a linear fit for early-type galaxies. The green line is a linear fit for late-type galaxies.

**ACKNOWLEDGEMENTS**

Yongyun Chen is grateful for financial support from the National Natural Science Foundation of China (no. 12203028). This work was support from the research project of Qujing Normal University (2105098001/094). This work is supported by the youth project of Yunnan Provincial Science and Technology Department (202101AU070146, 2103010006). Yongyun Chen is grateful for funding for the training Program for talents in Xingdian, Yunnan Province. QSGU is supported by the National Natural Science Foundation of China (12121003, 12192220, and 12192222). We also acknowledge the science research grants from the China Manned Space Project with no. CMS-CSST-2021-A05. This work was supported by the National Natural Science Foundation of China (11733001, U2031201, and 12433004).

**DATA AVAILABILITY**

All the data used here are available upon reasonable request. All data are in Table 1.

This paper has been typeset from a TeX/LaTeX file prepared by the author.